\documentclass[structabstract,rnote]{aa}  
\usepackage{graphicx}
\usepackage{txfonts}
\usepackage{rotating}
\usepackage{natbib}

\begin{document}
   \title{Discovery of a stellar companion to the nearby solar-analogue HD 104304
\thanks{Based on observations made with ESO Telescopes at the La Silla and Paranal Observatory under programme IDs 083.C-0145 and 084.C-0812, and on data obtained from the ESO Science Archive Facility.}
}

   \author{Carolin Schnupp\inst{1} \and
           Carolina Bergfors\inst{1} \and
           Wolfgang Brandner\inst{1} \and
           Sebastian Daemgen\inst{2} \and
           Debra Fischer \inst{3} \and
           Geoff Marcy\inst{4} \and
           Thomas Henning\inst{1}\and
           Stefan Hippler\inst{1}\and
           Markus Janson\inst{5}
}

   \institute{Max-Planck-Institut f\"ur Astronomie, K\"onigstuhl 17,
              69117 Heidelberg, Germany\\
              \email{schnupp@mpia.de}
\and European Southern Observatory, Karl-Schwarzschild-Stra\ss e 2, 85748 Garching, Germany
\and Department of Astronomy, Yale University, New Haven, CT 06520-8101, USA 
\and Department of Astronomy, University of California, Berkeley, CA 94720-3411, USA
\and Department of Astronomy, University of Toronto, 50 St George Street, Toronto, ON M5S 3H4, Canada}

   \date{Received ---; accepted ---}

 
   \abstract
   {Sun-like stars are promising candidates to host exoplanets and are often included in exoplanet surveys by radial velocity (RV) and direct imaging. In this paper we report on the detection of a stellar companion to the nearby solar-analogue star HD 104304, which previously was considered to host a planetary mass or brown dwarf companion.}
   {We searched for close stellar and substellar companions around extrasolar planet host stars with high angular resolution imaging to characterize planet formation environments. }
   {The detection of the stellar companion was achieved by high angular resolution measurements, using the ``Lucky Imaging'' technique at the ESO NTT 3.5m with the AstraLux Sur instrument. We combined the results with VLT/NACO archive data, where the companion could also be detected. The results were compared to precise RV measurements of HD 104304, obtained at the Lick and Keck observatories from 2001-2010. }
   {We confirmed common proper motion of the binary system.  A spectral type of M4V of the companion and a mass of 0.21\,M$_\odot$ was derived. 
   Due to comparison of the data with RV measurements of the unconfirmed planet candidate listed in the Extrasolar Planets Encyclopaedia, we suggest that the discovered companion is the origin of the RV trend and that the inclination of the orbit of $i\approx35\degr$ explains the relatively small RV signal.}
   {}


   \maketitle
%

\section{Introduction}

HD 104304 (LHS 5206, HIP 58576) is a high proper motion star at a distance of 12.91 $\pm$ 0.13\,pc \citep{perryman97}. Because of its proximity, brightness, and observability from both hemispheres, it has been the subject of numerous studies. It is a chromospherically inactive star with log R$'$(HK) = $-4.933$ \citep{gray06} and with an X-ray luminosity of L$_{\rm X}$ = $1.2 \times 10^{20}$\,W \citep{schmitt04}. It has a spectral type of G8IV to K0IV, log g = 4.35 to 4.15, and [Fe/H] = 0.22 to 0.18 \citep{gray06,cenarro07}, and is a thin disk member \citep{soubiran05}. With an age of $\approx$8.5\,Gyr and a mass of $\approx$1.01\,M$_\odot$ \citep{takeda07} it is considered a close analogue to the Sun at approximately twice its present age, and is a target of a high-precision Doppler survey aimed at identifying planetary companions to Sun-like stars \citep{Wright2004}. HD 104304 has also been suggested as a target to search for habitable planets with future space missions \citep{lopez05}.

In 2008 we started a high-angular resolution imaging survey of exoplanet hosts stars with the goal to identify close stellar and substellar companions. The first results of this survey were reported by \citet{daemgen09}. HD 104304 was included in the survey because of the identification of a potential substellar companion by \cite{Marcy2007}.

In the following we will use the designation HD 104304A for the primary star and HD 104304B for the companion. 

\section{Observations and data reduction}

\begin{figure*}[htb]
\centering
\includegraphics[width=\textwidth]{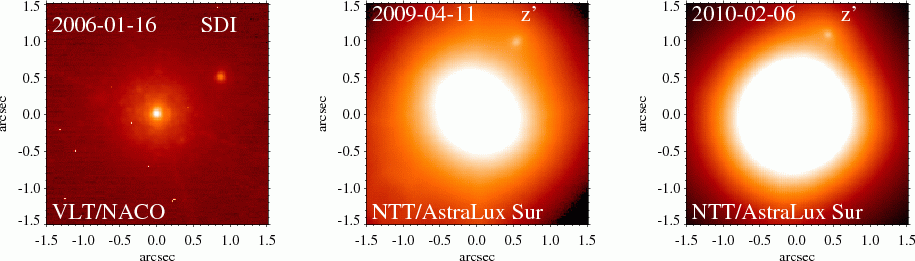}
\caption{AstraLux z$'$ and NACO/SDI images of the binary using a logarithmic
intensity scale. The field of view is $3'' \times 3''$, and North is up and East
 is to the left. The linear projected separation of the components is 12.8\,AU, 14.2\,AU and 14.8\,AU from the left to the right image. }
\label{fig1}
\end{figure*}

\subsection{AstraLux Sur}

Observations at the ESO 3.5m NTT were obtained with
the 'Lucky Imaging' instrument AstraLux Sur, a high-speed electron multiplying camera for Lucky Imaging observations \citep{hippler09}. The instrument is a clone of the common user instrument AstraLux Norte at the Calar Alto 2.2 m telescope \citep{hormuth08}. It provides a field of view of 16\,$\times$\,16\,arcseconds. 

HD 104304 was observed as part of a larger sample of (candidate) exoplanet host stars stars on April 11, 2009, and again on  February 6, 2010 in i$'$ and z$'$-band. Total integration times were 30\,s corresponding to the best 10\,\% of 10000 frames with an individual  t$_{\rm exp} = 0.030$s. Figure 1 shows the AstraLux images of the binary as well as the VLT/NACO data (see section 2.2).

The data were reduced by the AstraLux online data reduction pipeline as explained by \cite{hormuth08}.
Pixel scale and image orientation were determined by observations of star clusters with well-determined astrometry.

\subsection{VLT/NACO}

After the detection of the companion, we searched the VLT archive for high-angular
resolution observations of HD 104304.
VLT observations of HD 104304 using the  Spectral Differential Imager (SDI) mode of NACO were obtained on January 18, 2006 in service mode under the programme-ID 076.C-0762(A) for Damien S\'{e}gransan, and retrieved from the ESO archive. NACO/SDI\footnote{For more information on NACO/SDI see  http://www.eso.org/sci/facilities/paranal/instruments/naco/inst/} provides four simultaneous images in three different filters in the near infrared H-band. 
Total integration times were 20$\times$172.5\,s. The companion was clearly detected in the NACO images. 

\begin{figure}[htb]
\centering
\includegraphics[width=5cm]{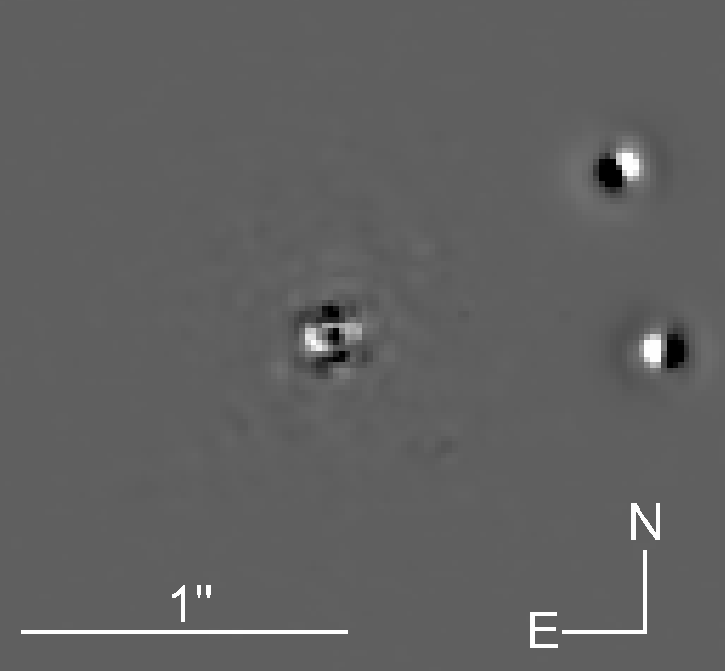}
\caption{With the SDI+ADI pipeline reduced NACO/SDI frames of HD 104304. In the center of the image the residuals of the subtracted primary star can be seen. The white and black dots on the right half side of the image are the companion on the different orientated images. }
\label{fig2}
\end{figure}

We also checked the deeper NACO/SDI frames for additional closer and fainter
companions by reducing the data with a dedicated SDI+ADI pipeline in IDL, see Figure \ref{fig2}.
For details on this pipeline see \cite{Janson2007}.
These data would enable us to detect a companion 7.5 mag fainter in H than HD
104304A with a detection limit of 3 sigma at an angular separation of 0.3\arcsec \ (as
assumed for the unconfirmed planet in http://exoplanet.eu).
No evidence for additional companions could be found.

\subsection{Determination of binary parameters}

Figure \ref{fig1}
shows a comparison of the AstraLux  z$'$, and the NACO  
images of the HD 104304 binary. In the NACO data set, we analysed the four acquisition images with the shortest
integration time to derive precise relative astrometry and photometry for both
components. For the analysis a pixel scale of 17.32\,mas/pixel was assumed. 
Aperture photometry was conducted to determine the brightness ratio. The position of companion and primary star were determined by fitting a Gaussian profile.

In the AstraLux data set, the procedure to determine the brightness ratio was slightly different, as the companion is situated inside the seeing halo of the parent star. Aperture photometry was conducted, taking also into account the halo background. This background was determined on the opposite site of the primary star from where the companion was situated, assuming a symmetric Point Spread Function (PSF). 
To determine position angle and separation, PSF subtraction was applied to the primary star. The PSF for subtraction was created by mirroring the half of the primary star without companion to the opposite side. On this PSF subtracted image, the position of the binary companion was determined by fitting a Gaussian profile. The position of the parent star was determined by fitting a Gaussian profile to the original data.  The results can be found in Table \ref{binprop}.

\begin{table*}[htb]
\centering
\caption[]{Binary properties}
         \label{binprop}
\begin{tabular}{lcccc}
\hline\hline
Date&Filter & separation & PA    & mag. diff.   \\
    &      & [$''$] & [deg] & [mag]  \\ \hline
2006-01-18  &NACO SDI     &$0.993 \pm 0.005$  & $300.1 \pm 0.1$  & $3.80 \pm 0.05 $  \\

2009-04-11$^1$   & SDSS i$'$   &$1.104 \pm 0.002$   & $333.02 \pm 0.08$  & $5.5 \pm 0.3$    \\

 &SDSS z$'$&&&$4.8 \pm 0.2$\\
2010-02-06$^{1,2}$    & SDSS z$'$  &$1.15 \pm 0.02$   & $340.45 \pm 0.15$  &    \\
\hline
\end{tabular}
\begin{quote}
 $^1$ Separation and position angle are the mean values of these parameters in i$'$ and z$'$-band. \newline
 $^2$ The magnitude difference of this data set was not determined as the seeing was much worse than in the 2009 data. 
\end{quote}

\end{table*}

\section{Physical properties of the HD 104304 binary}

\subsection{Common proper motion}
          
HD 104304 is a high proper-motion star with 
$\mu_{\rm RA} = 141.98 \pm  0.71$\,mas/yr and 
$\mu_{\rm DEC} = -483.34 \pm 0.42$\,mas/yr. 
In the  3.25 years that passed between the VLT/NACO and the first AstraLux observations, HD 104304 moved $461.44 \pm 2.3$\,mas to the East, and $1570.9 \pm 1.4$\,mas to the
South. In the same period, the separation between HD 104304A and B changed from $995 \pm 2$\,mas to $1104 \pm 2$\,mas, and the position angle increased from $300.19 \pm 0.01$\,deg to $333.02 \pm 0.08$\,deg (see 
Table \ref{binprop}). If the companion had been a background object, the corresponding numbers would have been a change in separation to $1404 \pm 2$
mas and in position angle to $314.47 \pm 0.03$ deg, which can be exluded
with high significance. Thus both sources form a physical binary, and the change in position angle can be attributed to orbital motion around the common center of mass of both components.

\subsection{Spectral types and component masses}

As stated by  \citet{kraus07}, the literature on intrinsic magnitudes and
basic astrophysical parameters of main-sequence stars comprises a rather
heterogeneous set. In order to minimize systematic errors in the determination of
the properties of the HD 104304 binary system, we base the following analysis
therefore on the homogenized set  of parameters as compiled and computed by \citet{kraus07}.
Assuming that the average of the
brightness differences in the three NACO/SDI spectral bands is close to the brightness
difference in H-band, we can derive the absolute H-band magnitudes of both
components based on the 2MASS photometry of the unresolved binary star. The absolute
H-band magnitude of HD 104304B derived this way is consistent with a  spectral type of
M4V.

To determine the i$'$ magnitude of the binary companion the magnitudes given by \citet{just08} of the unresolved HD 104304 binary were used. By a comparison of the magnitude with spectral types given by \citet{kraus07} the spectral type of the companion was derived. 
Component A has a spectral type of G8IV to K0IV which leads also to a spectral type of M4V of HD 104304B in i$'$-band. 
We note that for mid M-dwarfs,  i$'$-band magnitudes vary strongly with spectral type.
Thus an uncertainty by +0.4 mag in i$'$ corresponds to an uncertainty by less than
half a spectral-subclass. By interpolating the mass scale in Table 5 from \citet{kraus07} we derive a mass of $0.21^{+0.03}_{-0.02 }$\,M$_\odot$ for HD 104304B.

The derived absolute magnitudes of both components in H and i$'$-band are listed in Table \ref{phot}.

\begin{table}[htb]
\centering
\caption[]{Component Photometry (absolute magnitudes)}
         \label{phot}
\begin{tabular}{lcccc}
\hline\hline
Component& m$_{\rm H}$ &m$_{\rm i'}$ &SpT&mass    \\
     & [mag] & [mag]     \\ \hline
A    & $3.5 \pm 0.2$ & $4.6 \pm 0.3$& G8IV to K0IV& 1.01\,M$_\odot$\\
B    & $7.3 \pm 0.2$ & $10.1 \pm 0.4$&M4V& $0.21^{+0.03}_{-0.02 }$\,M$_\odot$ \\
\hline
\end{tabular}
        \end{table}

\subsection{Orbital parameters}

While the projected separation of the binary changed only by a relatively small
amount from 995\,mas to 1104\,mas, the position angle increased quite substantially by 32.83\,deg over a period of 3.25 years.
This gives some evidence that the orientation of the orbit is closer to
face-on than to edge-on (i.e.\ $i<$45$^\circ$).

The HIPPARCOS parallax of HD 104304 is $77.48 \pm 0.80$\,mas   	 
 \citep{perryman97}, corresponding
to a  distance of $12.91 \pm 0.13$\,pc. Hence the angular
separations correspond to projected separations of $\approx$12.8\,AU 
to $\approx$14.8\,AU.

Taking into account only the change of the position angle during the 3.25 years, assuming a circular orbit and
a semimajor axis of $\approx$14\,AU, we would expect an orbital
period of $\approx$36\,yr. With these assumed orbital parameters, Kepler's third law yields a system mass of $\approx$2.1\,M$_\odot$. Hence, the real eccentric orbit has a smaller semimajor axis or the orbital period is larger, as the system has a mass of $\approx$1.2\,M$_\odot$ (see Table \ref{phot}).

To  precisely determine the orbital parameters, more astrometric data points are necessary. Therefore astrometric monitoring of the binary over the next years is required.

\subsection{Comparison with data of unconfirmed planet candidate}

\begin{figure*}[htb]
\begin{center}
$\begin{array}{cc}
\includegraphics[width=9.0cm,angle=0]{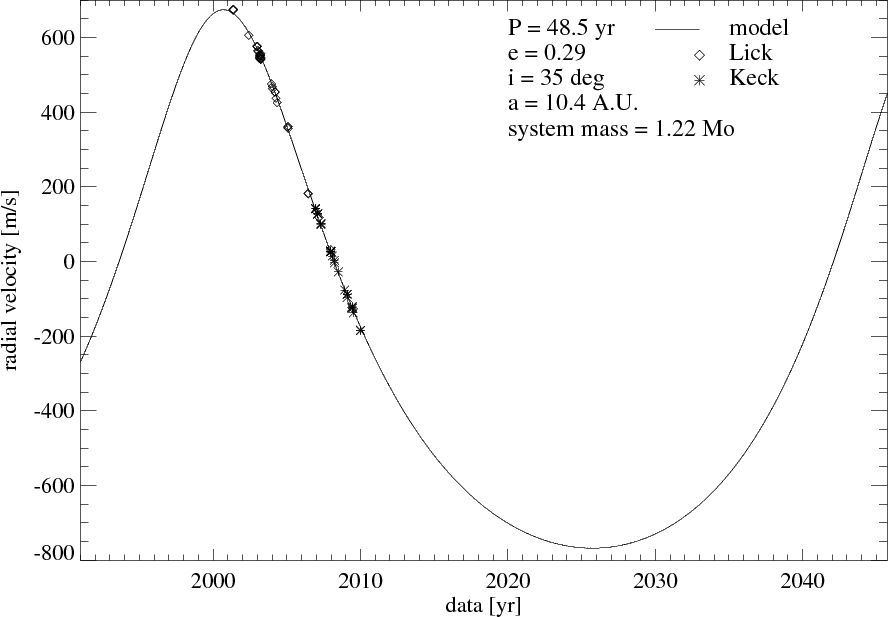} &
\includegraphics[width=8.8cm,angle=0]{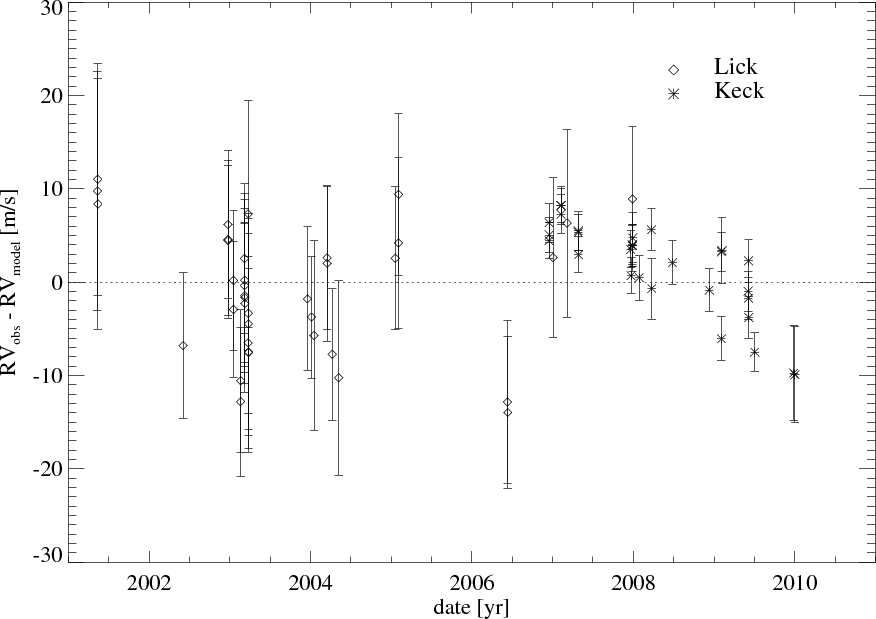}\\
\end{array}$
\end{center}
\caption[Theoretical RV curve for HD 104304]{Left: Theoretical RV curve of the binary system with a period of $P$ = 48.5 years, an inclination of $i=35\degr$ and an eccentricity of $e=0.29$. Triangles are the Lick RV measurements from 2001-2007 and asterisks are the Keck RV measurements from 2006-2010. Right: Residuals of the model. Diamonds are the residuals of the model with respect to the Lick measurements, asterisks with respect to the Keck measurements.} 
\label{rvcurve}
\end{figure*}

Based on its RV trend, HD 104304 has been identified as a candidate
host of a substellar object 
\citep{Marcy2007}. This RV trend was measured during 5 years, from 2001-2006. In these 5 years a change of $\approx$500\,m/s in the RV signal was detected.

We calculated the estimated radial velocity curves for a binary system with 1.22\,M$_{\odot}$, different orbital periods between $P$ = 35\,years and $P$ = 60\,years different inclinations from $i=5\,\degr$ to $i=50\,\degr$ and different eccentricities . 

To these theoretical RV curves, we compared precise RV measurements of HD 104304, taken over a period of 9 years. RV measurements at the Lick observatory have been obtained during a period of 6.5 years, from May 2001 to December 2007. At the Keck observatory, RV measurements have been taken over a period of 3 years, from December 2006 to January 2010.
As can be seen on Figure \ref{rvcurve}, the calculated RV curve for an inclination of $i=35\,\degr$, an eccentricity of $e=0.29$ and a period of $P$ = 48.5\,years compares well with the RV data. The identification of the unconfirmed planet by \cite{Marcy2007} was also based on these RV measurements from the Lick observatory, only on a shorter timescale than the RV data presented here. Therefore we suggest that the extrasolar planet candidate listed in the Extrasolar Planets Encyclopaedia\footnote{The Extrasolar Planets Encyclopaedia can be found at
http://exoplanet.eu and is maintained by Jean Schneider.} is identical with the stellar companion presented in this paper. 

While a plausible set of orbital parameters is derived, the analysis also hints that
thus far only around 20\% of the orbit has been covered by RV monitoring, and less
than 10\% of the orbit by astrometric observations. Thus further monitoring is
required in order to  derive a robust set of orbital parameters.

Theoretical calculations were made in order to check  if a real planet/brown dwarf companion orbiting HD 104304 would have  been visible in the NACO/SDI frames.

The reduced NACO/SDI data (see Figure \ref{fig2}) would enable us to detect a companion 7.5 mag fainter in H than HD
104304A at an angular separation of 0.3\arcsec, i.e. a source with an absolute magnitude m$_{\rm H}$ = 11.0 mag for a detection limit of 3 sigma. In
the age range 5 to 10 Gyr, according to Baraffe et al. (2003), this corresponds to a
star with 83 M$_{\rm Jup}$ (0.08 M$_\odot$).  Thus while we can exclude an
additional stellar companion with a mass $\ge$83 M$_{\rm Jup}$ at a projected
separation $\ge$3.9 AU, a substellar (brown dwarf or planetary mass companion) would
not be visible in the NACO images.

\section{Conclusions}

The combination of NACO/SDI archival data and AstraLux Sur data made it possible to confirm the stellar companion around HD 104304. Based on the photometry, a spectral type of M4V and a mass of 0.21 M$_\odot$ were derived for HD 104304B. 

The primary star is well studied, e.g. it has a known metallicity of [Fe/H] = 0.22 to 0.18 \citep{gray06,cenarro07}, an age of $\approx$\,8.5\,Gyr \citep{takeda07} and a mass nearly equal to that of our Sun. This knowledge is a valuable asset for the further characterisation of any stellar or substellar companion in this system.

By modeling the theoretical RV curve  and comparing it to the available RV observations obtained over a period of 9 years, we have shown that the observed RV trend could be explained by the detected stellar companion. A theoretical RV curve with an eccentricity of $e=0.29$, an orbital period of $P$ = 48.5\,years and an inclination of the orbit of $i=35\degr$~fits best the RV measurements. 
This gives evidence that the planet candidate listed in the Extrasolar Planets
Encyclopaedia around HD 104304 is identical with the presented stellar companion. 
Astrometric monitoring of the binary system over the next few years is necessary to confirm this statement and to precisely determine the orbit and its orientation. 

In the case of confirmation of a planet around HD 104304, this system would
be highly interesting in the context of planet formation and migration around 
close binaries with solar analogue host stars.

\begin{acknowledgements}
We are particulary grateful to Sam (Karl) Wagner, Armin Huber and Ralf-Rainer Rohloff
for their help in preparing and commissioning AstraLux Sur as well as all the
technical staff at ESO La Silla and MPIA involved in the project.
\end{acknowledgements}


\bibliographystyle{aa}	
\bibliography{14740}
\end{document}